\documentclass[11pt,a4paper]{article}

\usepackage{amssymb}
\usepackage{amsmath}
\usepackage{graphics}
\usepackage{epsfig}
\usepackage{float}
\usepackage{graphicx}
\usepackage{amsfonts}
\usepackage{amsthm}
\usepackage{newlfont}

\begin{document}

\title{\textbf{\large{Optimization of the Neutrino Oscillation Parameters \\using Differential Evolution}}}
\vspace{1cm}
\author{Ghulam Mustafa\thanks{g\_mustafa61@yahoo.com}, Faisal Akram\thanks{faisal.chep@pu.edu.pk}, Bilal Masud\thanks{bilalmasud.chep@pu.edu.pk} \\
\textit{Centre for High Energy Physics, University of the Punjab,
Lahore(54590), Pakistan.}}
\date{}
\maketitle

\begin{abstract}
We combine Differential Evolution, a new technique, with the
traditional grid based method for optimization of solar neutrino
oscillation parameters $\Delta m^2$ and $\tan^{2}\theta$ for the
case of two neutrinos. The Differential Evolution is a population
based stochastic algorithm for optimization of real valued
non-linear non-differentiable objective functions that has become
very popular during the last decade. We calculate well known
chi-square ($\chi^2$) function for neutrino oscillations for a grid
of the parameters using total event rates of chlorine (Homestake),
Gallax+GNO, SAGE, Superkamiokande and SNO detectors and
theoretically calculated event rates. We find minimum $\chi^2$
values in different regions of the parameter space. We explore
regions around these minima using Differential Evolution for the
fine tuning of the parameters allowing even those values of the
parameters which do not lie on any grid. We note as much as 4 times
decrease in $\chi^2$ value in the SMA region and even better
goodness-of-fit as compared to our grid-based results. All this
indicates a way out of the impasse faced due to CPU limitations of
the larger grid method. \vspace{.5cm}\noindent
\end{abstract}
\maketitle
\section{Introduction}
\qquad The flux of solar neutrino was first measured by Raymond
Davis Junior and John N. Bahcall at Homestake in late 1960s and a
deficit was detected between theory (Standard Solar Model) and
experiment \cite{Bahcall1969}. This deficit is known as the
\textit{Solar Neutrino Problem}. Several theoretical explanations
have been given to explain this deficit. One of these is neutrino
oscillations, the change of electron neutrinos to an other neutrino
flavour during their travel from a source point in the sun to the
detector at the earth surface~\cite{PRD17_1978}. There was no
experimental proof for the neutrino oscillations until 2002 when
Sudbury Neutrino Observatory (SNO) provided strong evidence for
neutrino oscillations~\cite{PRL89_2002}. The exact amount of
depletion, which may be caused by the neutrino oscillations,
however, depends upon the neutrino's mass-squared difference $\Delta
m^2\equiv m^2_2-m^2_1$ ($m_1$ and $m_2$ being mass eigen-states of
two neutrinos) and mixing angle $\theta$, which defines the relation
between flavor eigen-states and mass eigen-states of the neutrinos,
in the interval $[0,\pi/2]$.

The data from different neutrino experiments have provided the base
to explore the field of neutrino physics. In the global analysis of
solar neutrino data, we calculate theoretically expected event rates
with oscillations at different detector locations and combine it
with experimental event rates statistically through the chi-square
($\chi^2$) function, as defined below by Eq.(\ref{chi1}), for a grid
of values of the parameters $\mathrm{\Delta m^2}$ and $\mathrm{tan^2
\theta}$. The values of these parameters with minimum chi-square in
different regions of the parameter space suggest different
oscillation solutions. The names of these solutions, found in the
literature, along with specification of the regions in the parameter
space are: Small Mixing Angle (SMA: $10^{-4}\leq \mathrm{tan^2
\theta}\leq3 \times 10^{-2} ,\,\,
3\times10^{-7}\mathrm{eV^2}\leq\mathrm{\Delta
m^2}\leq10^{-4}\mathrm{eV^2}$), Large Mixing Angle (LMA:
$3\times10^{-2}\leq \mathrm{tan^2 \theta} \leq 2,\,\,
2\times10^{-6}\mathrm{eV^2}\leq\mathrm{\Delta
m^2}\leq10^{-3}\mathrm{eV^2}$), Low Probability Low Mass (LOW:
$3\times10^{-2}\leq \mathrm{tan^2 \theta} \leq 2,\,\,
10^{-8}\mathrm{eV^2}\leq\mathrm{\Delta m^2}\leq
2\times10^{-6}\mathrm{eV^2}$) and Vacuum Oscillation (VO: $0.1\leq
\mathrm{tan^2 \theta} \leq 1,\,\,
10^{-11}\mathrm{eV^2}\leq\mathrm{\Delta
m^2}\leq10^{-8}\mathrm{eV^2}$) \cite{AP17_2002}. Extensive work has
been done on the global analysis of solar neutrino data
\cite{PRD58_1998,NPB573_2000,NPBPS91_2001,PRD63_2001,JHEP05_2001,JHEP08_2001,PRD65_2002,JHEP07_2002,PRD66_2002,PRD67_2003,NPBPS143_2005}
and now is the era of precision measurement of the neutrino
oscillation parameters \cite{CPL26_3_2009,CPL26_8_2009}.

Traditionally, the whole parameter space ($10^{-4}\leq \mathrm{tan^2
\theta} \leq 10$,  $10^{-13}\mathrm{eV^2}\leq\mathrm{\Delta
m^2}\leq10^{-3}\mathrm{eV^2}$) is divided into a grid of points by
assigning a variable to each parameter and varying its
logarithm uniformly. The chi-square values are calculated for each point in
the parameter space either by using $\mathrm{^8B}$ flux constrained
by the Standard Solar Model, e.g., BS05(OP)~\cite{APJ621_2005} in
our case, or by using unconstrained $\mathrm{^8B}$ flux
\cite{JHEP05_2001} where it is varied about the value predicted by
the Standard Solar Model. The global minimum chi-square value
$\chi^2_{min}$ is found and $\mathrm{100\,\beta}\%$ C.L. (Confidence
Level) contours are drawn in the $\mathrm{tan^2
\theta}-\mathrm{\Delta m^2}$ plane by joining points with
$\chi^2=\chi^2_{min}+\Delta \chi^2$ for different confidence levels.
From the chi-square distribution one can easily find that $\Delta
\chi^2=2.28, 4.61, 5.99, 9.21, 11.83$ for 68\%, 90\%, 95\%, 99\% and
99.73\% C.L. for two degrees of freedom. Minimum chi-square values
are found in all the regions and the goodness-of-fit, corresponding
to each of the minimum chi-square, is calculated. To find the each
goodness-of-fit the chi-square distribution is used and confidence
level $\mathrm{100(1-\beta)} \%$, corresponding to the minimum
chi-square in the region and the degree of freedom of the analysis,
is calculated~\cite{AP17_2002,JHEP05_2001}. In our analysis we used
total event rates of chlorine (Homestake), Gallax+GNO, SAGE,
Superkamiokande, SNO CC and NC experiments. So the number of degrees
of freedom was 4 (6(rates)--2(parameters: $\mathrm{tan^2 \theta}\,\,
\mathrm{and}\,\, \mathrm{\Delta m^2}$)).

When we use the Differential Evolution (DE), the parameters are
randomly selected in the given range and checked for a decrease of
chisquare, in contrast with the traditional grid based method as
described in the above paragraph. Thus we selected the vectors with
least chi-square values, in different regions of the selected grid,
as starting points and used DE for the fine tuning of the parameters
by exploring region around the selected vectors in the parameter
space.

Here in section \ref{chisquare}, we define the chi-square ($\chi^2$)
function for the solar neutrino oscillations. We use the same
$\chi^2$ function definition in the algorithm of DE as well as in
the traditional method. In section \ref{DEA}, we describe algorithm
of Differential Evolution along with its salient features. In
section \ref{analysis} and \ref{opt}, we describe results of global
analysis by grid and those obtained using Differential Evolution
respectively. Our conclusions are given in section \ref{conclusion}.
\section{Chi-square ($\chi^2$) Function Definition \label{chisquare}}
\qquad In our $\chi^2$ analysis, we used the updated data of total
event rates of different solar neutrino experiments. We followed the
$\chi^2$ definition of ref.~\cite{Astropart3_1995} and included
chlorine (Homestake)~\cite{APJ496_1998}, weighted average of Gallax
and GNO~\cite{NPBPS110_2002}, SAGE~\cite{PRC80_2009},
Superkamiokande~\cite{PRD83_2011}, SNO CC and SNO
NC~\cite{PRL101_2008} total rates. The expression for the $\chi^2$
is given as:
\begin{equation}
\chi^2_{\mathrm{Rates}} = \sum_{j_1,j_2=1,6} (R^{th}_{j_1} -
R^{exp}_{j_1})[V_{j_1j_2}]^{-2}(R^{th}_{j_2} - R^{exp}_{j_2})
\label{chi1},
\end{equation}
where $R^{th}_{j}$ is the theoretically calculated event rate with
oscillations at detector $j$ and $R^{exp}_{j}$ is the measured rate.
For chlorine, Gallax+GNO and SAGE experiments  $R^{th}$ and
$R^{exp}$ are in the units of SNU (1 SNU=$10^{-36}$
captures/atom/sec) and for Superkamiokande, SNO CC and SNO NC these
are used as ratio to SSM Eq.(\ref{chi6}) below. $V_{j_1j_2}$ is the
error matrix that contains experimental (systematic and statistical)
errors and theoretical uncertainties that affect solar neutrino
fluxes and interaction cross sections. For the calculation of the
error matrix $V_{j_1j_2}$ we followed ref.~\cite{Astropart3_1995}
and for updated uncertainties we used ref.~\cite{PRD62_2001}. For
the calculation of theoretical event rates, using
Eqs.(\ref{chi2}-\ref{chi5}) below, we first found the time average
survival probabilities, over the whole year, of electron neutrino
$\langle P_{ee}^k(E_\nu)\rangle$ ($E_\nu$ is the neutrino energy in
MeV) at the detector locations for the $k^{th}$ neutrino source and
for the grid of 101 $\times$ 101 values of $\mathrm{\frac{\Delta
m^2}{E}}$ and $\mathrm{tan^2 \theta}$ following the prescriptions
described in ref.~\cite{JHEP05_2001}. For the uniform grid interval
distribution we used the parameters $\mathrm{\frac{\Delta m^2}{E}}$
and $\mathrm{tan^2 \theta}$ as exponential functions of the
variables $x_1$ and $x_2$ as:
\begin{equation}
\mathrm{\frac{\Delta m^2}{E}} = 10^{(0.1 x_1-13)}\label{chi1A}
\end{equation}
and
\begin{equation}
\mathrm{tan^2 \theta}= 10^{-2(2-0.025 x_2)}\label{chi1B}
\end{equation}
so that discrete values of $x_1$ and $x_2$ from 0 to 100 cover the
entire $\mathrm{tan^2 \theta}-\mathrm{\Delta m^2}$ parameter space.
We used the expression for the average expected event rate in the
presence of oscillation in case of Chlorine and Gallium detectors
given as:
\begin{equation}
R_j^{th} = \sum_{k=1\, \mathrm{to}\, 8} \phi_k
\int^{E_{max}}_{E^j_{th}} dE_\nu
\lambda_k(E_\nu)[\sigma_{e,j}(E_\nu)\langle P_{ee}^k(E_\nu)\rangle].
\label{chi2}
\end{equation}
Here $E^j_{th}$ is the process threshold for the $j$th detector
($j$=1,2,3 for Homestake, Gallax+GNO and SAGE respectively). The
values of energy threshold $E^j_{th}$ for Cl, Ga detectors are
0.814, 0.233 MeV respectively \cite{RMP75_2003}. $\phi_k$ are the
total neutrino fluxes taken from BS05(OP)~\cite{APJ621_2005}. For
Gallium detector all fluxes contribute whereas for Chlorine detector
all fluxes except pp flux contribute. $\lambda_k(E_\nu)$ are
normalized solar neutrino energy spectra for different neutrino
sources from the sun, taken from refs. \cite{PRC54_1996,RMP60_1988},
and $\sigma_{e,j}$ is the interaction cross section for $\nu_e$ in
the $j$th detector. Numerical data of energy dependent neutrino
cross sections for chlorine and gallium experiments is available
from ref.~\cite{PRC54_1996}. Event rates of Chlorine
\cite{APJ496_1998} and Gallium \cite{NPBPS110_2002,PRC80_2009}
experiments and those calculated from Eq.(\ref{chi2}) directly come
in the units of SNU.

Superkamiokande and SNO detectors are sensitive for higher energies,
so $\phi_k$ are the total $\mathrm{^8B}$ and hep fluxes for these
detectors respectively. The expression of the average expected event
rate with oscillations for elastic scattering at SK detector is as
below:
\begin{equation}
N_{SK}^{th} = \sum_{k=1,2} \phi_k \int^{E_{max}}_{0} dE_\nu
\lambda_k(E_\nu)\times\{\sigma_{e,j}(E_\nu)\langle
P_{ee}^k(E_\nu)\rangle+\sigma_{\mu,j}(E_\nu)[1-\langle
P_{ee}^k(E_\nu)\rangle]\}. \label{chi3}
\end{equation}
Here $\sigma_e$ and $\sigma_\mu$ are elastic scattering cross
sections for electron and muon neutrinos that we took from ref.
\cite{PRD51_1995}.

For the SNO charged-current (CC) reaction, $\nu_e d \rightarrow
e^-pp$, we calculated event rate using the expression:
\begin{equation}
N_{CC}^{th} = \sum_{k=1,2} \phi_k \int dE_\nu
\lambda_k(E_\nu)\sigma_{CC}(E_\nu)\times \langle P_{ee}^k(E_\nu)
\rangle. \label{chi4}
\end{equation}
Here $\sigma_{CC}$ is $\nu d$ CC cross section of which
calculational method and updated numerical results are given in
refs. \cite{PRC63_2001} and ~\cite{NPA707_2002} respectively.

The expression for the SNO neutral-current (NC) reaction, $\nu_x d
\rightarrow \nu_x \,p \,n \, (x=e,\, \mu, \, \tau)$, event rate is
given as:
\begin{equation}
N_{NC}^{th} = \sum_{k=1,2} \phi_k \int dE_\nu
\lambda_k(E_\nu)\sigma_{NC}(E_\nu)\times (\langle
P_{ee}^k(E_\nu)\rangle+ \langle P_{ea}^k(E_\nu)\rangle).
\label{chi5}
\end{equation}
Here $\sigma_{NC}$ is $\nu d$ NC cross section and $\langle
P_{ea}^k(E_\nu)\rangle$ is the time average probability of
oscillation into any other active neutrino. We used updated version
of CC and NC cross section data from the website given in ref.
\cite{NPA707_2002}. In case of oscillation of the $\nu_e$ into
active neutrino only, $\langle P_{ee}^k(E_\nu)\rangle+\langle
P_{ea}^k(E_\nu)\rangle=1$ and $N_{NC}^{th}$ is a constant.

For Superkamiokande \cite{PRD83_2011} and SNO \cite{PRL101_2008}
experiments, the event rates come in the unit of $10^6
\mathrm{cm^{-2}s^{-1}}$. We converted these rates into ratios to SSM
predicted rate. We also calculated theoretical event rates as ratios
to SSM predicted rate in order to cancel out all energy independent
efficiencies and normalizations \cite{PRD63_2001}.
\begin{equation}
R_{j}^{th} = \frac{N_j^{th}}{N_j^{SSM}} \label{chi6}
\end{equation}
Here $N_j^{SSM}$ ($j$=4,5,6 for SK, SNO CC and SNO NC respectively)
is the predicted number of events assuming no oscillations. We used
the Standard Solar Model BS05(OP)~\cite{APJ621_2005} in our
calculations. Theoretical event rates, so calculated, were used in
Eq.(\ref{chi1}) to calculate the chi-square function for different
points in the $\mathrm{tan^2 \theta}-\mathrm{\Delta m^2}$ parameter
space.

\section{Differential Evolution \label{DEA}}
\qquad Differential Evolution (DE) is a simple population based,
stochastic direct search method for optimization of real valued,
non-linear, non-differentiable objective functions. It was first
introduced by Storn and Price in 1997 \cite{JOGO_1997}. Differential
Evolution proved itself to be the fastest evolutionary algorithm
when participated in First International IEEE Competition on
Evolutionary Optimization \cite{ICEC_1996}. DE performed better when
compared to other optimization methods like Annealed Nelder and Mead
strategy \cite{CUP_1992}, Adaptive Simulated Annealing
\cite{JMCM_1993}, Genetic Algorithms \cite{Goldberg_1989} and
Evolution Strategies \cite{Schefel_1995} with regard to number of
function evaluations (nfe) required to find the global minima. DE
algorithm is easy to use, robust and gives consistent convergence to
the global minimum in consecutive independent trials
\cite{JOGO_1997,Omega_2005}.

The general algorithm of DE \cite{Springer_1998} for minimizing an
objective function carries out a number of steps. Here we summarize
the steps we carried out for minimizing the $\chi^2$ function
defined in section \ref{chisquare}. We did optimization of the
$\chi^2$ function individually for different regions of the
parameter space to do one fine tuning in each region. The results of
the optimization are reported in the section \ref{opt} below.
\subsection*{Step I}
An array of vectors was initialized to define a population of size
\texttt{NP}=20 with \texttt{D}=2 parameters as
\begin{equation}
\mathrm{\textbf{x}}_i=\textit{x}_{j,i} \,\,\, \mathrm{where}
\,\,\,\, i=1,2,.....,\texttt{NP} \,\,\, \mathrm{and} \,\,\,
j=1,..,\texttt{D}. \label{DE1}
\end{equation}
The parameters, involved here, are $x_1$ and $x_2$ of
Eqs.(\ref{chi1A}) and (\ref{chi1B}) on which $\mathrm{\Delta m^2/E}$
and $\mathrm{tan^2\theta}$ depend. Upper and lower bounds ($b_{j,U}$
and $b_{j,L}$),  individually for different regions of the parameter
space described in the introduction section, for the $x$ values were
specified and each vector $i$ was assigned a value according to
\begin{equation}
\textit{x}_{j,i}=\mathrm{rand}_j(0,1)\cdot (b_{j,U}-b_{j,L})+b_{j,L}
\label{DE2}
\end{equation}
where $\mathrm{rand}_j \in[0,1]$ is $j^{th}$ evaluation of a uniform
random number generator. The $\chi^2$ function was calculated for
each vector of the population and the vector with least $\chi^2$
function value was selected as base vector $\textbf{x}_{r_o}$.
\subsection*{Step II}
Weighted difference of two randomly selected vectors from the
population was added to the base vector $\textbf{x}_{r_o}$ to
produce a mutant vector population $\textbf{v}_{i}$ of \texttt{NP}
trial vectors. The process is known as \textit{mutation}.
\begin{equation}
\textbf{v}_{i}=\textbf{x}_{r_o}+F\cdot(\textbf{x}_{r_1}-\textbf{x}_{r_2}).
\label{DE3}
\end{equation}
Here the scale factor $F\in[0,2]$ is a real number that controls the
amplification of the differential variation. The indices
$r_1,r_2\in[1,\texttt{NP}]$ are randomly chosen integers and are
different from $r_o$.

Different variants of DE \textit{mutation} are denoted by the
notation `\texttt{DE/x/y/z}', where \texttt{x} specifies the vector
to be mutated which can be ``rand" (a randomly chosen vector) or
``best" (the vector of the lowest $\chi^2$ from the current
population), \texttt{y} is the number of difference vectors used and
\texttt{z} is the crossover scheme. The above mentioned variant
Eq.(\ref{DE3}) is \texttt{DE/best/1/bin}, where the best member of
the current population is perturbed with \texttt{y}=1 and the scheme
\texttt{bin} indicates that the crossover is controlled by a series
of independent binomial experiments. The two variants, reported in
the literature \cite{JOGO_1997,Omega_2005}, very useful for their
good convergence properties, are \texttt{DE/rand/1/bin}
\begin{equation}
\textbf{v}_{i}=\textbf{x}_{r_1}+F\cdot(\textbf{x}_{r_2}-\textbf{x}_{r_3}),
\label{DE6}
\end{equation}
and \texttt{DE/best/2/bin}
\begin{equation}
\textbf{v}_{i}=\textbf{x}_{ro}+F\cdot(\textbf{x}_{r_1}+\textbf{x}_{r_2}-\textbf{x}_{r_3}-\textbf{x}_{r_4})\,.
\label{DE7}
\end{equation}

For our problem, we used the variant \texttt{DE/best/2/bin}
Eq.(\ref{DE7}) for DE mutation, where 2 difference vectors were
added to the base vector. The values of $F$ we used are reported in
section \ref{opt} below.

\subsection*{Step III}
The parameters of \textit{mutant vector} population Eq.(\ref{DE7})
were mixed with the parameters of \textit{target vectors}
Eq.(\ref{DE1}) in a process called uniform \textit{crossover} or
discrete recombination. After the cross over the \textit{trial
vector} became:
\begin{equation}
\textbf{u}_{i}=\textit{u}_{j,i}= \left\lbrace \begin{array}{rr}
              \textit{v}_{j,i} & \mathrm{If}\,\,(\mathrm{rand}_j(0,1) \leq \texttt{Cr}\,\, \mathrm{or}\,\, j=j_\mathrm{rand}),\\
              \textit{x}_{j,i} &  \mathrm{otherwise}. \hspace{3.9cm}  \end{array} \right. \label{DE4}
\end{equation}
Here $\texttt{Cr}\in[0,1]$ is the cross over probability that
controls fraction of the parameters inherited from the mutant
population (the values of \texttt{Cr} we used are given in section
5), $\mathrm{rand}_j \in[0,1]$ is the output of a random number
generator and $j_\mathrm{rand}$ $\in [1,2]$ is a randomly chosen
index.
\subsection*{Step IV}
The $\chi^2$ function was evaluated for each of the trial vector
$\textbf{u}_{i}$ obtained from Eq.(\ref{DE4}). If the trial vector
resulted in lower objective function than that of the target vector
$\textbf{x}_{i}$, it replaced the target vector in the following
generation. Otherwise the target vector was retained. (This
operation is called \textit{selection}.) Thus the target vector for
the next generation became:
\begin{equation}
\textbf{x}^\prime_i= \left \lbrace \begin{array}{rr}
              \textbf{u}_{i} & \mathrm{If}\,\,\chi^2(\textbf{u}_{i}) \leq \chi^2(\textbf{x}_{i}), \hspace{1.0cm} \\
              \textbf{x}_{i} &  \mathrm{otherwise}. \hspace{2.0cm}  \end{array} \right.
\label{DE5}
\end{equation}

The processes of mutation, crossover and selection were repeated
until the optimum was achieved or the number of iterations
(generations) specified in section \ref{opt} were completed.

\section{Analysis from the Selected Grid \label{analysis}} 
\begin{figure}[!t]
\begin{center}
\includegraphics[angle=0,width=0.50\textwidth]{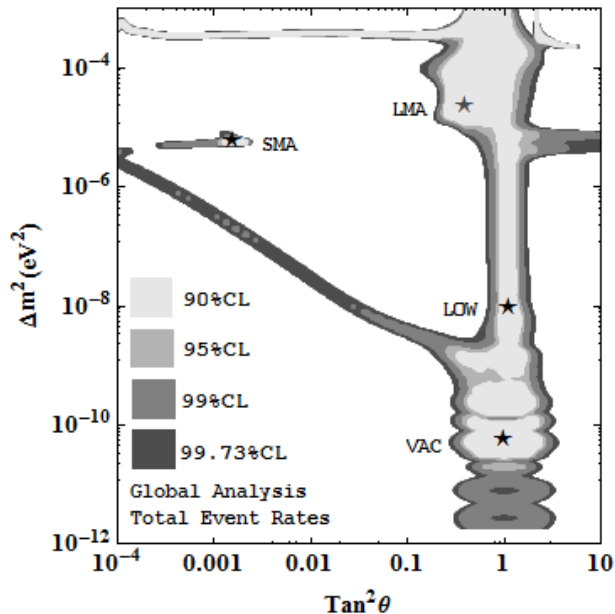}
\caption{Our global
solutions for the total rates. The input data
includes total event rates of chlorine  \cite{APJ496_1998}, weighted
average of Gallax and GNO \cite{NPBPS110_2002}, SAGE
\cite{PRC80_2009}, Superkamiokande \cite{PRD83_2011}, SNO CC and SNO
NC \cite{PRL101_2008}. The increasing grey level shows 90\%, 95\%,
99\%, 99.73\% C.L. Our best-fit points in different regions are
marked by stars.} \label{ga}
\end{center}
\end{figure}

\qquad Figure \ref{ga}
and Table \ref{parameters} show our best fit oscillation parameters,
in different regions, calculated using a grid of 101 $\times$ 101
points of the parameter space. The symbol of star shows the best fit
points in the respective regions of the parameter space.
Calculations of goodness-of-fit and confidence level are described
in the introduction section. We used chi-square function definition
of section \ref{chisquare}. We used $\mathrm{^8B}$ flux constrained
by the Standard Solar Model BS05(OP). We saw that the point with
global minimum or the best fit point in the parameter space lies in
the LMA region with $\Delta \mathrm{m^2}=2.512 \cdot 10^{-5}
\mathrm{eV^2}$ and $\tan^2 \theta=3.981 \cdot 10^{-1}$ that is
consistent with the results found in the literature where SNO data
is included in the analysis
\cite{JHEP05_2001,JHEP08_2001,JHEP07_2002}. Before including the SNO
data the best fit was found in the SMA region \cite{RMP75_2003}.

A selection of a fine grid with larger number of points in the
parameter space, of course, will give better results but limitations
of the CPU time restricted us, like others, to a grid with a small
number of points. But we point out that even without increasing the
number of points in the grid we can get lower $\chi^2$ and better
g.o.f. by \textit{fine tuning} of the oscillation parameters using
DE. We describe what we mean by \textit{fine tuning} and report our
improvements obtained this way in the next section.

\begin{table}[!t]
\begin{center}
\begin{tabular}{ccccc}
\hline \hline
Solution & $\Delta \mathrm{m^2}(\mathrm{eV^2})$ & $\tan^2 \theta$ &  $\mathrm{\chi^2_{min}}$ &g.o.f.\\
\hline
LMA  & $2.512 \cdot 10^{-5}$ & $3.981 \cdot 10^{-1}$ & 0.808 & 93.77\%\\
VAC & $6.31 \cdot 10^{-11}$ & $1.00 \cdot 10^{0}$ & 1.268 & 86.72\%\\
LOW & $1.00 \cdot 10^{-8}$ & $1.122 \cdot 10^0$ & 4.09 & 39.46\%\\
SMA & $6.31 \cdot 10^{-6}$ & $1.585 \cdot 10^{-3}$ & 7.78 & 10.01\%\\
\hline \hline
\end{tabular}
\end{center}
\caption{Our best-fit values of the oscillation parameters $\Delta
\mathrm{m^2}$,  $\tan^2 \theta$ along with  $\mathrm{\chi^2_{min}}$
(4 \,d.o.f) (6(rates)-2(parameters: $\mathrm{tan^2 \theta},
\mathrm{\Delta m^2}$)) and g.o.f. corresponding to Figure \ref{ga}.}
\label{parameters}
\end{table}

\section{Optimization of the Chi-square Function using DE \label{opt}}
\qquad We have described algorithm of the Differential Evolution in
detail in section \ref{DEA}. We wrote the subroutine of the
chi-square function, denoted by $\chi^2$, following the definition
of chi-square in section \ref{chisquare}, that depends on $x_1$ and
$x_2$ and used it as objective function of the DE algorithm. We
combined the traditional grid-based method with DE in two aspects:
First, we used the survival probabilities $\langle
P_{ee}^k(E_\nu)\rangle$ already calculated for the discrete values
of $x_1$ and $x_2$ for our grid of 101 $\times$ 101 points of the
parameter space and interpolated them to the continuous values of
$x_1$ and $x_2$ to calculate event rates and chi-square function in
DE algorithm. We used cubic polynomial fit for the interpolation
purpose to fit the data. Second, we used the points with minimum
chi-square in different regions of the selected grid Table
\ref{parameters} as the starting points (and members of the
respective population array) and explored the space around them for
the \textit{fine tuning}. That is, we searched for the points with
smaller $\chi^2$ values and better goodness-of-fit of the
oscillation parameters.

\begin{table}[!t]
\begin{center}
\begin{tabular}{lllllll}
\hline \hline
Solution &Iterations &$\Delta \mathrm{m^2}(\mathrm{eV^2})$ & $\tan^2 \theta$ & $\mathrm{\chi^2_{min}}$ &$\mathrm{\chi^2_{best}}$ &g.o.f.\\
\hline
 &1-7 & $2.51189\cdot10^{-5}$&$3.98107\cdot10^{-1}$&0.808314&&93.77\%\\
 &8-9& $2.50999 \cdot 10^{-5}$ & $3.97855 \cdot 10^{-1}$ & 0.807316 & &\\
 &10-13& $2.40927 \cdot 10^{-5}$ & $3.97684 \cdot 10^{-1}$ & 0.804711 & &\\
 &14& $2.49798 \cdot 10^{-5}$ & $3.97287 \cdot 10^{-1}$ & 0.804564 & &\\
 LMA&15& $2.49307 \cdot 10^{-5}$ & $3.97028 \cdot 10^{-1}$ & 0.804289 & &\\
 &16& $2.45884 \cdot 10^{-5}$ & $3.97912 \cdot 10^{-1}$ & 0.804424 & &\\
 &17-21& $2.46633 \cdot 10^{-5}$ & $3.9751 \cdot 10^{-1}$ & 0.803315 & &\\
 &22-29& $2.43264 \cdot 10^{-5}$ & $3.98097 \cdot 10^{-1}$ & 0.803192 & &\\
 &30-50& $2.45084 \cdot 10^{-5}$ & $3.9751 \cdot 10^{-1}$ & 0.802953 &0.802953 &93.82\%\\
\hline
 &1& $6.30957 \cdot 10^{-11}$ & $1.0$ & 1.26779 & &86.72\%\\
 &2-3& $6.64977 \cdot 10^{-11}$ & $1.0277$ & 1.260239 & &\\
 VAC&4-6& $6.70641 \cdot 10^{-11}$ & $1.01912$ & 1.25977 & &\\
 &7-43& $6.6423 \cdot 10^{-11}$ & $0.993134$ & 1.25948 & &\\
 &44-45& $6.6423 \cdot 10^{-11}$ & $0.998353$ & 1.25945 & &\\
 &46-50& $6.70041 \cdot 10^{-11}$ & $0.99326$ & 1.25939 &1.25939 &86.82\%\\
\hline
 &1-3 & $6.30957 \cdot 10^{-6}$ & $1.58489 \cdot 10^{-3}$ & 7.77933 & &10.01\%\\
 SMA&4 & $6.19532 \cdot 10^{-6}$ & $1.64599 \cdot 10^{-3}$ & 6.24974 & &\\
 &5 & $6.10563 \cdot 10^{-6}$ & $1.48276 \cdot 10^{-3}$ & 5.89341 & &\\
 &6-50 & $5.48095 \cdot 10^{-6}$ & $1.72371 \cdot 10^{-3}$ & 1.86456 &1.86456 &75.97\%\\
\hline
 &1 & $1.0 \cdot 10^{-8}$ & $1.12208$ & 4.18897 & &39.46\%\\
 &2-4 & $2.37807 \cdot 10^{-8}$ & $1.03198$ & 3.98339 & &\\
 &5-9 & $2.95404 \cdot 10^{-8}$ & $1.03198$ & 3.97624 & &\\
 LOW&10 & $3.3042 \cdot 10^{-8}$ & $1.03069$ & 3.97605 & &\\
 &11 & $2.80796 \cdot 10^{-8}$ & $1.02741$ & 3.9728 & &\\
 &12-20 & $3.17357 \cdot 10^{-8}$ & $1.02741$ & 3.96267 & &\\
 &21-50 & $3.14543 \cdot 10^{-8}$ & $1.02723$ & 3.96125 &3.96125 &41.12\%\\
\hline \hline
\end{tabular}
\end{center}
\caption{The results of the oscillation parameters during different
iterations of the DE algorithm. The improved values of the
oscillation parameters $\Delta \mathrm{m^2}$, $\tan^2 \theta$ along
with $\mathrm{\chi^2_{best}}$ (4 d.o.f) and g.o.f. using
Differential Evolution strategy \texttt{DE/best/2/bin} corresponding
to Figure \ref{de1} are presented. Note in the $1^{st}$ row of
different regions, the points with minimum chi-square given in table
\ref{parameters} are taken as first members of the population
arrays. The other members of the arrays, for different regions, are
selected randomly using DE.}\label{parametersDE1}
\end{table}

\begin{figure}[!h]
\begin{center}
\begin{tabular}{p{8.0cm}p{8.0cm}p{0.0cm}p{0.0cm}}
\includegraphics[angle=0,width=0.45\textwidth]{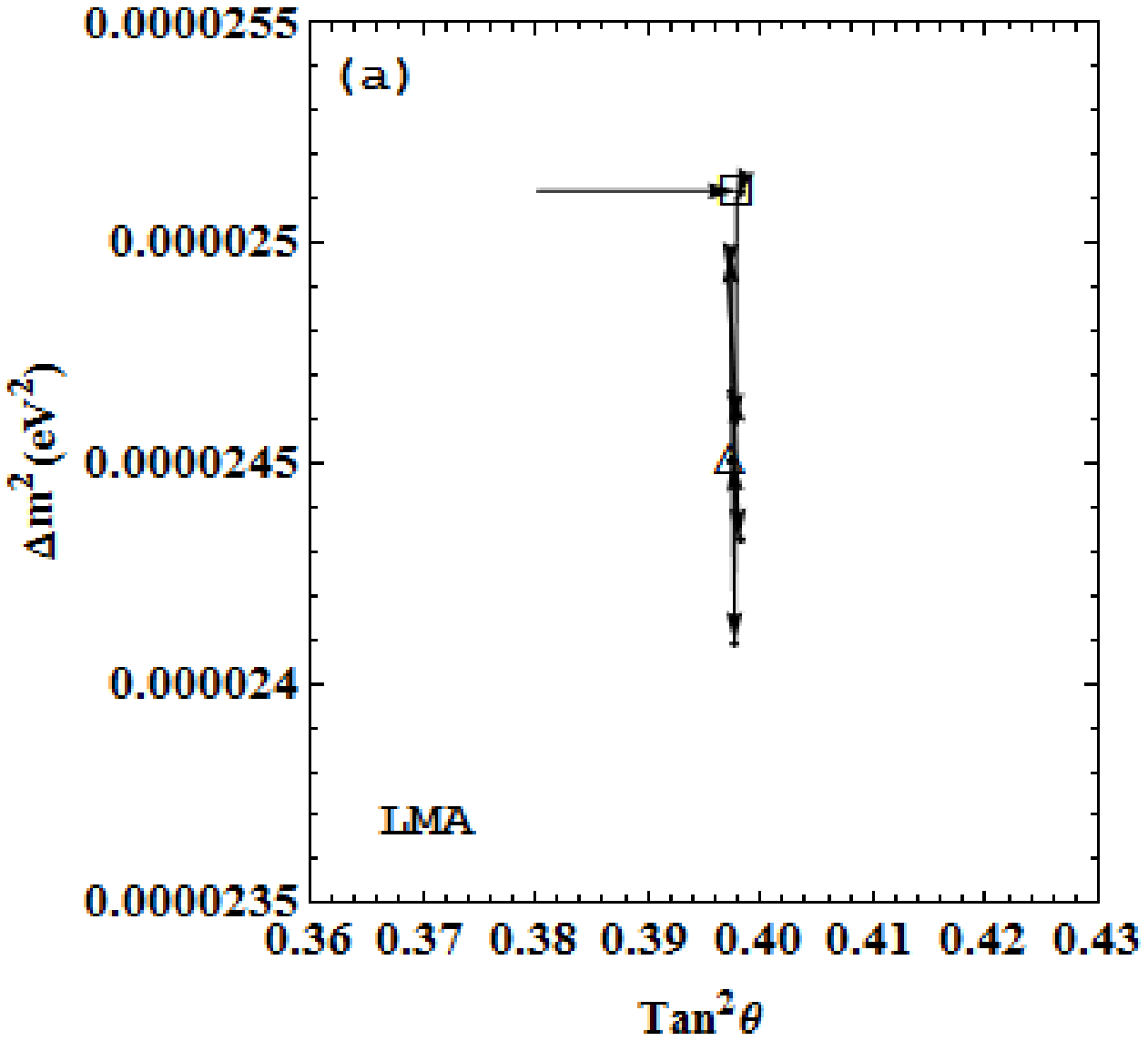} &
\includegraphics[angle=0,width=0.45\textwidth]{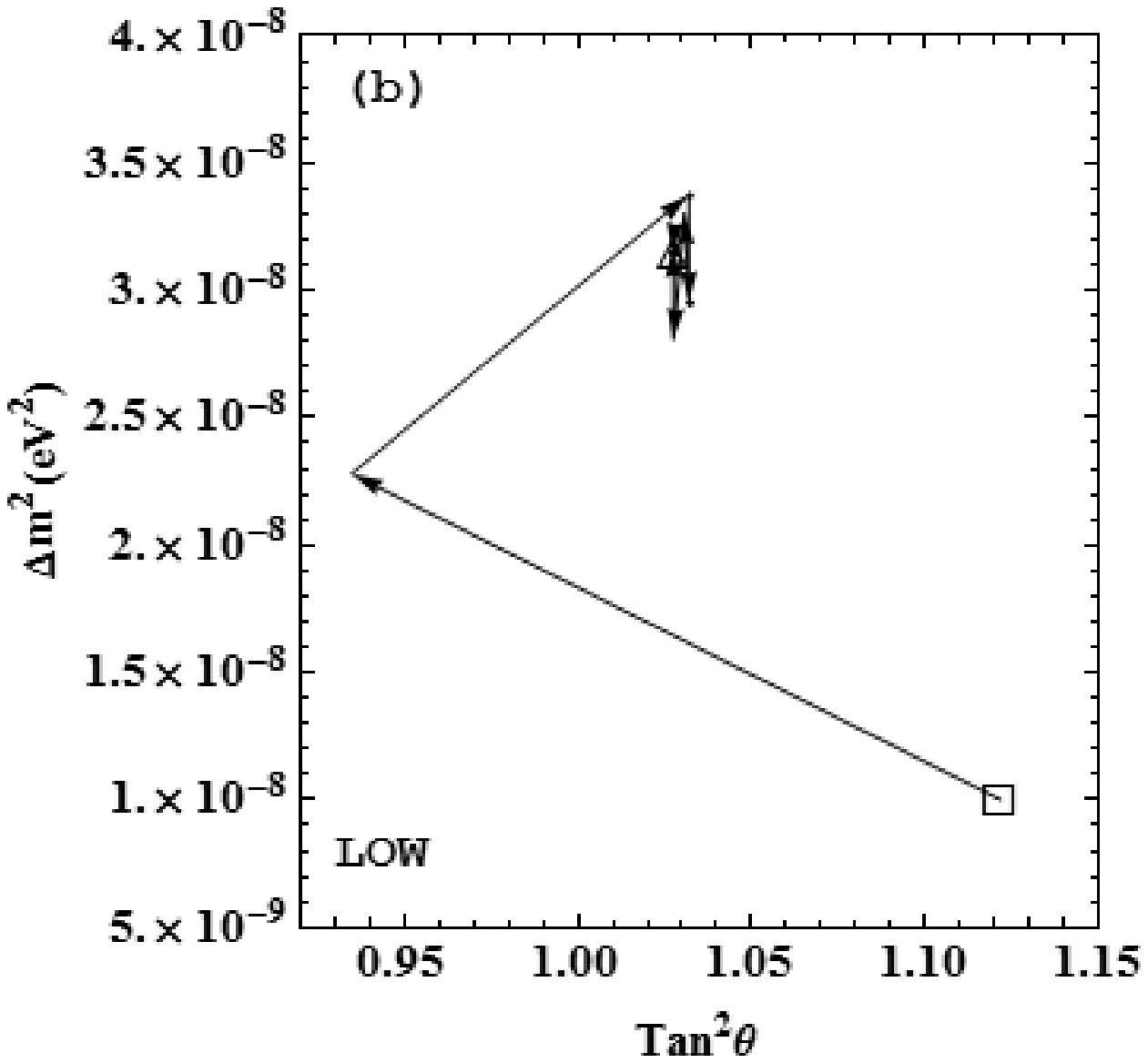} &\\
\includegraphics[angle=0,width=0.45\textwidth]{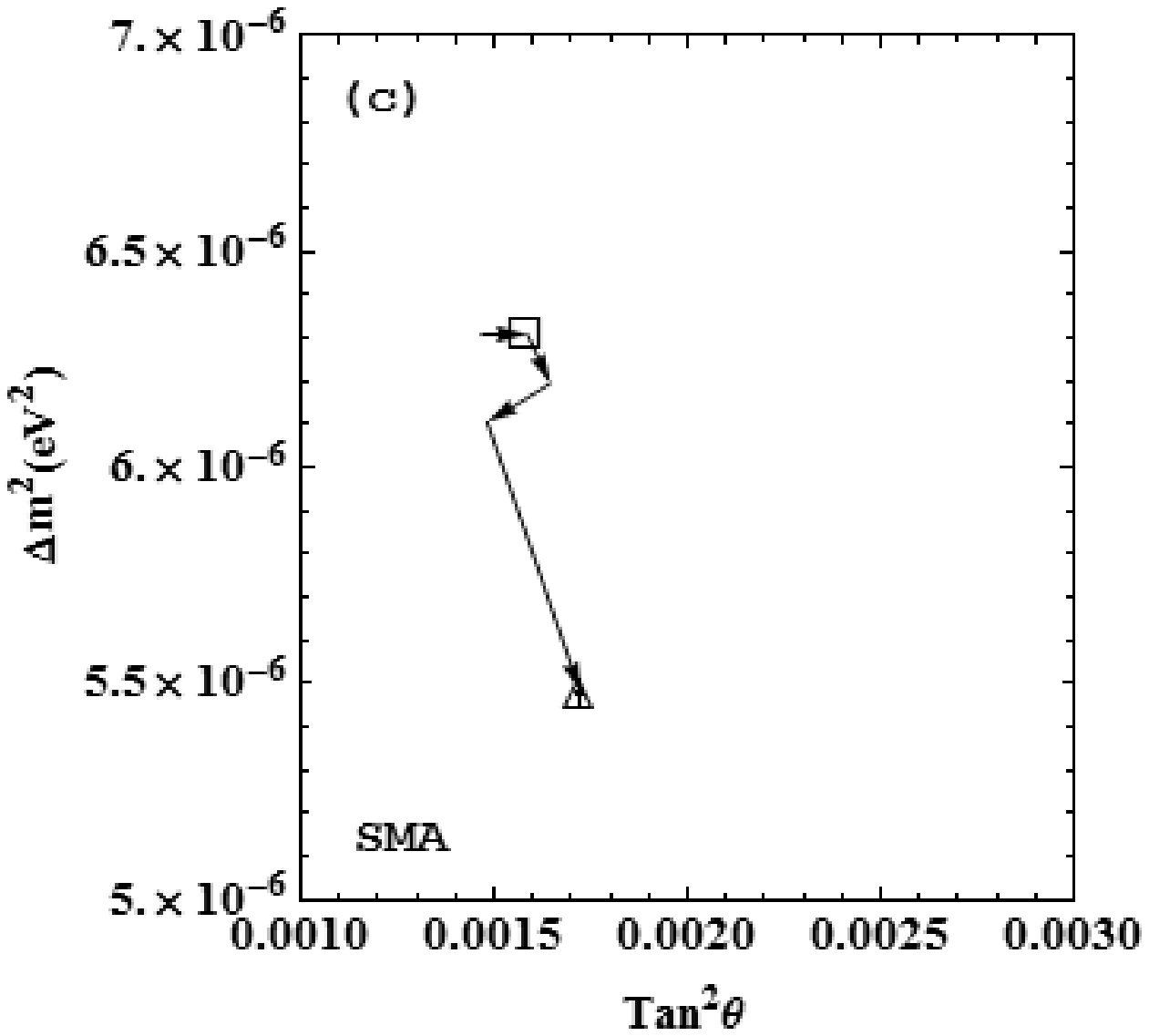} &
\includegraphics[angle=0,width=0.45\textwidth]{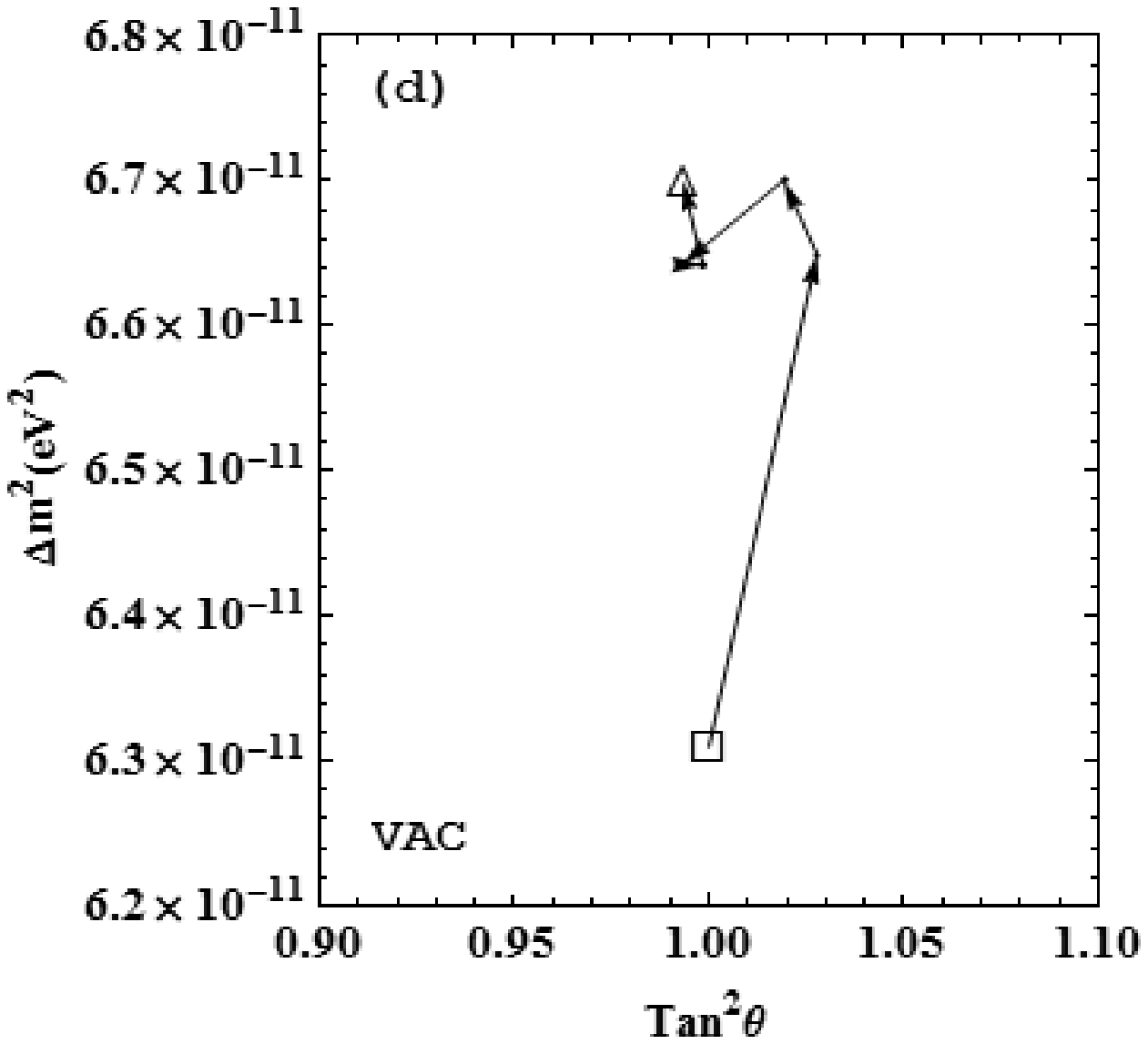}
\end{tabular}
\caption{Track of the DE algorithm for optima in different regions
using the strategy \texttt{DE/best/2/bin}. The square symbol shows
the best point of the 101 $\times$ 101 grid and triangle symbol
shows the best point after fine tuning using DE.} \label{de1}
\end{center}
\end{figure}

In our analysis, the values of DE control variables \texttt{F} and
\texttt{CR} were taken as 0.4 and 0.9 respectively for the LMA, SMA
and VAC regions. For the LOW region \texttt{F} and \texttt{CR} were
both taken as 0.3 for better convergence. Maximum number of
iterations were taken to be 50 for all regions. We took the best
point in a region of the 101 $\times$ 101 grid in Table
\ref{parameters} as the first member of the population in the first
iteration and used the strategy \texttt{DE/best/2/bin} for DE
mutation in all the remaining iterations/generations. The steps of
DE algorithm, described in section \ref{DEA}, are repeated for the
number of iterations specified.

Table \ref{parametersDE1} and Figure \ref{de1} show the results in
different regions during and after fine tuning of the oscillation
parameters using Differential Evolution. The value of
$\mathrm{\chi^2_{min}}$ persisted, rejecting all the mutations, for
the iterations mentioned in column 2 of Table \ref{parametersDE1}.
Accepted mutations resulted in new vectors whose components are
given in column 3 and 4 of the following rows.
$\mathrm{\chi^2_{best}}$ is the minimum chi-square value we obtained
in the region specified. In comparison to the results of Table
\ref{ga}, we note here as much as 4 times decrease in the
$\chi^2_{\mathrm{min}}$ of the SMA region after fine tuning using DE
along with a small decrease in all the other regions. Different
vectors in Figure \ref{de1} show the track of DE algorithm for
optima in different regions during iterations specified in Table
\ref{parametersDE1}.

\section{Conclusions \label{conclusion}}
\qquad Fine tuning of the neutrino oscillation parameters using
Differential Evolution has been introduced as a solution to the
impasse faced due to CPU limitations of the larger grid alternative.
We can explore the parameter space deeply due to real nature of the
parameters $x_1$ and $x_2$ using DE in contrast to discrete nature
of these parameters in the traditional grid based method. We
conclude that combination of Differential Evolution along with
traditional method provides smaller chi-square values and better
goodness-of-fit of the neutrino oscillation parameters in different
regions of the parameter space. We also note a significant change in
the results of $\chi^2_{\mathrm{min}}$ and g.o.f. in the SMA region
after the fine tuning using DE. Even though it is a local decrease,
it indicates importance of the exploration of the points within the
grid and the efficiency that can be achieved through DE.

\section*{Acknowledgements}
We are thankful to the Higher Education Commission (HEC) of Pakistan
for its financial support through Grant No.17-5-2(Ps2-044)
HEC/Sch/2004.

\end{document}